\documentstyle[12pt,epsf,epsfig,here]{article}    % Simple LaTeX, use with DINA4 as follows
\newlength{\dinwidth}                       
\newlength{\dinmargin}                      
\def\lsim{\mathrel{\rlap{\lower4pt\hbox{\hskip1pt$\sim$}}
    \raise1pt\hbox{$<$}}}                % less than or approx. symbol
\def\gsim{\mathrel{\rlap{\lower4pt\hbox{\hskip1pt$\sim$}}
    \raise1pt\hbox{$>$}}}                % greater than or approx. symbol
\newcommand{\beq}  {\begin{equation}}
\newcommand{\eeq}  {\end{equation}}
\newcommand{\bmath}{\begin{eqnarray}}
\newcommand{\emath}{\end{eqnarray}}
\newcommand{\TS}   {\textstyle}

\begin{document}
\begin{titlepage}
\begin{flushleft}
{\tt DESY 97-249 }  \\

{\tt hep-ph/9801300}
\end{flushleft}
\vspace*{3.cm}

\begin{center}  \begin{Large} \begin{bf}
A Study of the Polarized Structure Function {\boldmath {$g_{1}^{\rm p}(x,Q^{2})$}} and the
 Polarized Gluon Distribution {\boldmath $\Delta g(x,Q^{2})$}
at HERA\\
  \end{bf}  \end{Large}

  \vspace*{2cm}
  \begin{large}
A.~De Roeck$^{a,b}$, A.~Deshpande$^{c}$,~V.W.~Hughes$^{c}$,\\
~J.~Lichtenstadt$^{d}$, G.~R\"adel$^{b}$\\
 \end{large}
%\end{center}

\vspace*{.8cm}
$^a$ Deutsches Elektronen-Synchroton DESY, Notkestr.\ 85, 22603 Hamburg,
 Germany\\
$^b$ CERN, Div.\ PPE, 1211 Gen\`eve 23, Switzerland \\
$^c$ Department of Physics, Yale University, New Haven, CT 06520, 
USA \\
$^d$ School of Physics and Astronomy, The Raymond and Beverly Sackler Faculty of Exact\\ ~~Sciences,
Tel Aviv University, Tel Aviv 69978,  Israel \\

\vspace*{2.cm}
\begin{quotation}
\noindent
{\bf Abstract:}
We present estimates of possible data on spin-dependent asymmetries 
in inclusive scattering of high energy polarized electrons by 
high energy polarized protons at HERA with their statistical
errors and discuss systematic errors.
We show that these data will 
provide important information on the low-$x$ behavior of the polarized 
structure function $g_1$, and will reduce
the uncertainty in the determination of the first moment of the polarized
gluon distribution $\Delta g(x,Q^2)$ obtained from the
 QCD analysis  of $g_1$ in NLO. Furthermore, 
using asymmetries for di-jet events
from a polarized HERA would substantially reduce
the uncertainty in the shape of  $\Delta g(x,Q^2)$. Use of the information
on $\Delta g(x,Q^2)$ from the di-jet analysis 
in conjunction with the NLO QCD analysis of $g_{1}$ will provide an
accurate determination of $\Delta g(x,Q^2)$ and its first moment.
\end{quotation}
\vfill
\cleardoublepage
\end{center}
\end{titlepage}
%%%
%%%   SECTION 1
\section{Introduction}
Measurements of nucleon structure functions
by lepton-nucleon inclusive electromagnetic deep inelastic scattering (DIS) were  of fundamental importance in studying  nucleon structure 
and have provided crucial information for 
the development of perturbative QCD (pQCD).
The history of these experiments over the past 40 years has shown that
important new information has been obtained when measurements were extended
to new kinematic regions. In the mid-1950's at Stanford,  measurements
of elastic electron-proton scattering were extended to a higher $Q^{2}$ 
range of 1 ${\rm (GeV/c)}^{2}$ and for the first time it was  observed that 
the proton has a finite size~\cite{hof}.
In the late 1960's at SLAC  the extension of measurements
of inelastic inclusive electron scattering to the deep inelastic region of
$Q^{2} > 1$ (GeV/c)$^{2}$ led to the discovery of the parton substructure 
of the proton~\cite{FKT}. This substructure was further studied in muon-nucleon
deep inelastic scattering~\cite{EMC}. The extension of the data to
lower $x$ and higher $Q^2$ made possible by the HERA $ep$-collider showed
a surprisingly steep rise towards low $x$ of  the unpolarized 
structure function $F_2$~\cite{F2hera} whose explanation has led to 
an improved understanding of perturbative QCD.
Furthermore, these data as well as data from di-jet production~\cite{HERAG} 
enabled a precise determination of the gluon density in the nucleon.
 
  Polarized DIS, which measures the spin dependent structure function
of the nucleon has a similar history. 
The first measurements by the
Yale-SLAC collaboration \cite{Y-SLAC} extended down to $x=0.1$ and were
consistent with the naive quark-parton model
view that the nucleon spin is carried by its constituent quarks.  
These experiments were followed
up at CERN by the EM Collaboration~\cite{EMCpol}  extending the measurements
 to lower $x$ of 0.01. The results
 showed that the Ellis-Jaffe sum rule is violated, which within the
quark parton model implied that the contribution of the quark spins 
$\Delta \Sigma$ to the proton spin is small.
This surprising result stimulated a large amount of 
experimental
and theoretical work on polarized  structure functions~\cite{spinrev}.
The next generation of
experiments by the SM Collaboration at CERN~\cite{SMC,SMCp96}, by 
the SLAC  collaborations~\cite{E142,E143,E154}, and by the HERMES collaboration
at DESY~\cite{herm} 
reduced significantly the statistical and systematic uncertainties on the 
measurements and the kinematic range was extended to $x=0.003$.
In addition, spin structure functions for
the deuteron and neutron as well as the proton 
were measured,
thereby providing the first experimental verification of the 
fundamental model-independent Bjorken sum-rule.
On the theory side, in the QCD quark parton model,
 the breaking of the Ellis-Jaffe sum rule
can be  interpreted  as due to a positively polarized gluon distribution and/or
a negatively polarized strange quark contribution.
Hence precise information on the polarized gluon distribution is important
for the understanding of the spin structure of hadrons, and will be studied in 
this paper.

Despite the recent experimental progress, the 
polarized structure functions are still 
 measured in a  limited $x$ and $Q^2$ range and a large 
uncertainty on the first moment of $g_1$
comes from the unknown low-$x$ behavior of $g_1$. 
The kinematic range accessible by present-day  fixed
target experiments is limited and therefore 
to obtain a significant extension of the kinematic range the scattering 
of polarized electrons by polarized protons must be studied in a
collider mode such as HERA.
While the polarized quark distributions $\Delta q$ are accessible
in fixed target experiments from measurements of $g_1$,
the polarized gluon distribution $\Delta g$ 
is deduced  from scaling violations of $g_1$.
However, the good quality data on $g_1$ in the present kinematic range
 allows us to perform
complete next-to-leading order (NLO) QCD analyses of the $Q^2$
dependence of  $g_1$. Recently several such analyses 
have  been performed~\cite{BFRa,GRSV,SG,E154qcd} and have provided the 
first information on the 
first moment   of the polarized gluon distribution, 
$\int_0^1 \Delta g(x){\rm d}x$. 
These analyses suggest a fairly large contribution of the
gluon to the spin of the proton~\cite{SMCp96,ABFR}.
Nevertheless the uncertainties are  very large and the shape
of $\Delta g(x)$ is poorly constrained.
This analysis would benefit greatly from additional 
data over a wider kinematic range.
%It is therefore extremely important to devise measurements to reduce
%the uncertainty on the polarized gluon distribution.
 
\begin{figure}
        \epsfxsize=10cm
        \epsfysize=10cm
        \hfil
        \epsffile[20 20 410 410]{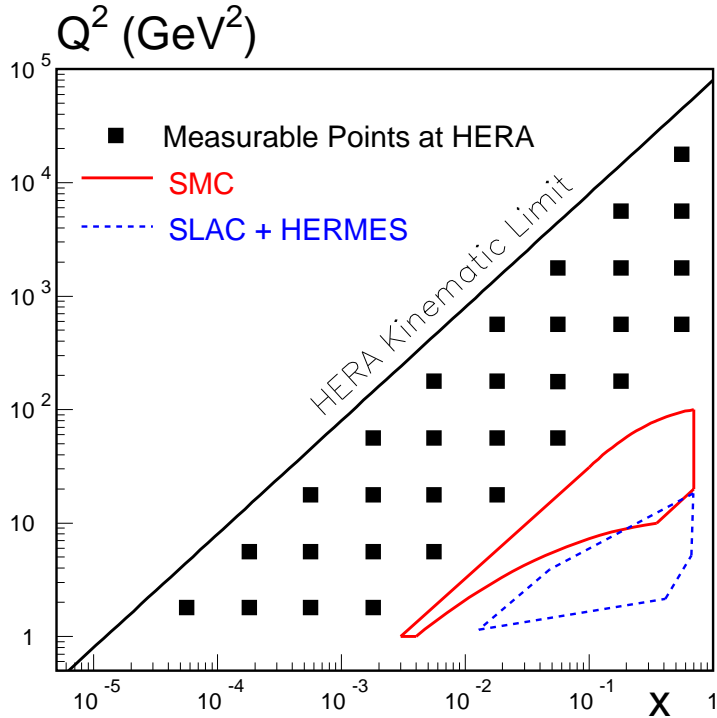}
        \hfil
        \caption{ Measurable $x-Q^{2}$ region with polarized HERA shown 
        with the presently explored regions by CERN(SMC), SLAC and 
        DESY(HERMES) experiments, and the kinematic limit of measurability 
        at HERA.}
\label{xq2_hera}
\end{figure}

The HERA collider at DESY provides 27.5~GeV electrons and
820 GeV protons. At present the 
proton beam at HERA is unpolarized whereas the 
electron/positron beam has a natural transverse polarization, 
which can be rotated into a longitudinal polarization in the 
interaction regions.
Once the protons in the HERA ring are polarized it will be possible 
to measure polarization asymmetries and thus explore spin
dependent structure functions in the full HERA kinematic range. 
The possible extension in the measured $x-Q^{2}$ range for spin structure 
function measurements is shown in Fig.~\ref{xq2_hera}.
The scientific and technical problems involved in obtaining
polarized protons in the HERA ring have been discussed in~\cite{barber}. 
The measurability of the polarized structure functions at HERA
has first been discussed in~\cite{HERAg1,g1old,abhay}.

%%%
%%%  Figure 1
%%%
 
In this paper we study the potential 
impact of a polarized HERA on the knowledge 
 of the polarized structure function
 and the polarized gluon distribution in the proton.
We first  show that  $g_1$ data from polarized colliding beam 
experiments at 
HERA would provide information  on the low-$x$ behavior of $g_1$ and
through  QCD analyses of $g_1$ would 
substantially improve the determination of the 
first moment of the polarized gluon distribution.
%We further  show  that  information can be obtained on the  
%shape of the polarized gluon distribution from the asymmetry of di-jet
%production cross-sections.
We present for the first time  an analysis which includes $\Delta g(x,Q^2)$
measurements from di-jet asymmetries together with the QCD analysis of
$g_{1}$ data from fixed target experiments and from HERA.
The determination of the polarized gluon distribution
$\Delta g(x,Q^{2})$ and its first moment improves substantially when using
such a combined analysis.

%%%%%%%%%%%%%%%%%%%%%%%%%%%%%%%%%%%%%%%%%%%%%%%%%%%%%%%%%%%%%%%%%%%%%%%%%%%%%%%%%%%
%%%   SECTION 2
%%%%%%%%%%%%%%%%%%%%%%%%%%%%%%%%%%%%%%%%%%%%%%%%%%%%%%%%%%%%%%%%%%%%%%%%%%%%%%%%%%%
%\section{Current status of $g_1(x,Q^2)$, $\Delta g(x,Q^{2})$ and di-jet events}
\section{Current status of polarized parton distributions}
\label{sec-g1stat}
The structure function 
$g_1$ is related to the polarized quark and gluon distributions through
\bmath
   g_1(x,t) & = & {\TS\frac{1}{2}} \langle e^2\rangle
        \int_x^1 \frac{{\rm d}y}{y} \Bigl [ 
        C_q^{\rm S}({\TS\frac{x}{y}},\alpha_s(t)) \Delta\Sigma(y,t)
      \nonumber \\
      & + & 2 n_f C_g({\TS\frac{x}{y}},\alpha_s(t)) \Delta g(y,t) 
    + C_q^{\rm NS}({\TS\frac{x}{y}},\alpha_s(t)) \Delta q^{\rm NS}(y,t) \Bigr],
\label{qcd:g1}
\emath
where $\langle e^2\rangle = n_f^{-1}\sum_{k=1}^{n_f} e_k^2$, 
$t = \ln(Q^2/\Lambda^2)$, 
$\Delta\Sigma$ and $\Delta q^{\rm NS}$ are the singlet and non-singlet
polarized quark distributions:
$$
\Delta\Sigma(x,t) =  \sum_{i=1}^{n_f} \Delta q_i(x,t),
\qquad
\Delta q^{\rm NS}(x,t)  =  \sum_{i=1}^{n_f} 
                ( e_i^2/\langle e^2\rangle -1)\Delta q_i(x,t),
\label{qcd:qsns}
$$
and $C_q^{\rm {S,NS}}(\alpha_s(Q^2))$ and $C_g(\alpha_s(Q^2))$ are 
the quark and gluon coefficient functions.

The $x$ and $Q^2$ evolution  of the polarized quark and gluon distributions
is given by the
DGLAP evolution equations~\cite{AP}.
% Altarelli-Parisi equations~\cite{AP}.
The full set of  coefficient functions~\cite{cf} and splitting 
functions~\cite{Pij} has been computed up to next-to-leading order in $\alpha_{s}$. 
%As in any perturbative calculation, 
At NLO 
splitting functions, coefficient functions
and parton distributions depend on the renormalization
and factorization scheme.
The scheme choice is arbitrary and parton distributions in
different factorization schemes are related to each other by well-defined 
linear transformations. 

\begin{figure}[hp]
        \epsfxsize=10cm
        \epsfysize=10cm
        \hfil \epsffile[0 0 520 540]{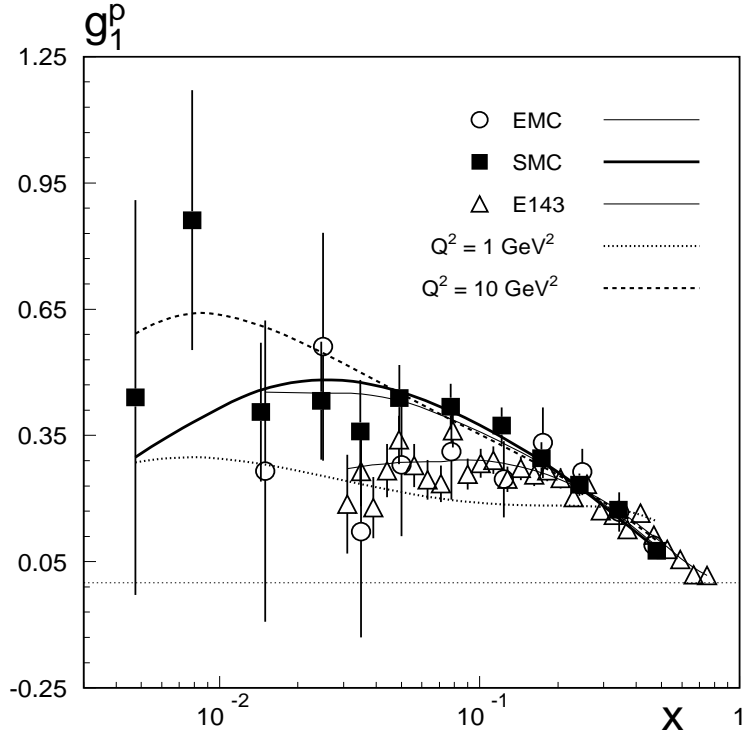} \hfil
        \caption { The NLO fit to proton $g_{1}^{\rm p}$ 
 %       , deuteron $g_1^{\rm d}$, and neutron $g_{1}^{\rm n}$ 
        data. The solid lines are the fit to the data at  measured $Q^{2}$ 
        values, and the dashed and the dotted lines are the fit evolved to 
        $Q^{2}=$ 1 and 10 GeV$^{2}$ respectively.}
        \label{qcdfit}
\end{figure}

The 
polarized parton distributions can be obtained by fitting the experimental  
spin structure 
function data. 
%The $Q^2$ dependence of the polarized quark and gluon
%distributions is determined by the Altarelli-Parisi equations~\cite{AP}. 
The distributions are parametrized at an initial scale 
and are evolved using the DGLAP  equations to values 
of $x$ and $Q^2$ of the data, 
varying the parameters to fit the experimental values of $g_1$~\cite{BFRa,GRSV,SG}. 
Here we follow the procedure used in
ref.~\cite{BFRa}. The initial parametrization at
$Q^2=1$~GeV$^2$ is assumed to have the form 
\begin{equation}
\label{part_dist}
\Delta f(x,Q^{2}) = 
N(\alpha_f,\beta_f,a_f)~\eta_f~x^{\alpha_f}(1-x)^{\beta_f}(1+a_f~x),
\end{equation}
where $N(\alpha,\beta,a)$ is fixed for each distribution
 by the normalization condition,
$N(\alpha,\beta,a)\int_{0}^{1} 
 x^{\alpha}(1-x)^{\beta}(1+ax){\rm d}x  =1$, and $\Delta
f$ denotes $\Delta \Sigma$, $\Delta q^{\rm NS}$, or $\Delta g$. 
With this normalization the parameters
$\eta_{g},\eta_{NS}$, and $\eta_{\Sigma}$ are 
the first moments of the polarized gluon distribution, the
non-singlet quark distribution (in the proton or in the neutron), 
and the singlet quark distribution,
respectively, at the initial scale.
In this parametrization $\alpha_f$ and $\beta_f$ determine the low-$x$ and
high-$x$ behavior of the parton distribution, respectively. The polynomial
describes the medium-$x$ region and was included  only for the parametrization
 of the singlet quark distribution.
 Some variations of the initial parton distributions
were also tried in order to check the dependence of the results on
the choice of the parametrization. The evolution was performed within
the AB factorization scheme 
(in which  $C_g^1=-{\alpha_s\over 4\pi}$). The strong coupling constant
$\alpha_s(M_Z^2)$ was taken to be $0.118\pm 0.003$.
The factorization scale $M^2$ and the renormalization scale $\mu^2$
were taken to be $M^2=k_1\cdot Q^2$ and $\mu^2=k_2 \cdot Q^2$, with 
$k_1=k_2=1$.
Further theoretical details of the fit procedure and analysis can be found in 
refs.~\cite{BFRa, ABFR}. 
Following this scheme, the SMC performed a  QCD analysis in NLO
to existing polarized structure function data for proton, neutron and
deuteron measured at SLAC, CERN and
 DESY~\cite{EMCpol,SMC,SMCp96,E142,E143,E154,herm}.
We repeated this fit and show 
the results for the best 
fit coefficients in Table 1. The world data
on $g_1^{\rm p}$ are compared to the best fit in Fig.~\ref{qcdfit}.
\begin{table}[hp]
\hfil
\begin{tabular}{||c|p{1cm}rcl|p{0.2cm}rcl||}
\hline\hline
Parameter & \multicolumn{4}{c|}{Fit 
to~\cite{EMCpol,SMC,SMCp96,E142,E143,E154,herm}} & \multicolumn{4}{c||}{HERA
 ${\cal L} = 500$ pb$^{-1}$} \\
\hline\hline
$\eta_g $ & & ~0.9 & $\pm$&  0.3 & & 0.9 &$\pm$& 0.2\\
$\eta_q $ & & ~0.40& $\pm$& 0.04  & &  0.38 &$\pm$& 0.03\\
$\eta_{NS}^{\rm p(n)}$ & & \multicolumn{3}{l|}{ $+(-) \frac{3}{4}g_A + \frac{1}{4}a_8$} &  &\multicolumn{3}{l||}{
 $+(-) \frac{3}{4}g_A  + \frac{1}{4}a_8$ }\\
\hline
$\alpha_{g}$  & & $-$0.5 &$\pm$& 0.3 & &  $-$0.6 &$\pm$& 0.1 \\
$\alpha_{q}$  & &  ~0.7 &$\pm$& 0.3 & & 0.9 &$\pm$& 0.3\\
$\alpha_{NS}^{\rm p}$ & & $-$0.1 &$\pm$& 0.1 & &  $-$0.1 &$\pm$& 0.1\\
$\alpha_{NS}^{\rm n}$ & & $-$0.1 &$\pm$& 0.2 & & 0.1 &$\pm$& 0.2\\
\hline

$\beta_{g}$   & &  \multicolumn{3}{l|}{~~~~4.0 (fixed)}& & \multicolumn{3}{c||}{  4.0 (fixed) } \\
$\beta_{q}$   & &  ~1.3 &$\pm$& 0.7  & & 1.5 &$\pm$& 0.7\\
$\beta_{NS}^{\rm p}$  & &   ~1.5 &$\pm$& 0.3 & & 1.4 &$\pm $& 0.3\\
$\beta_{NS}^{\rm n}$  & &   ~2.6 &$\pm$& 0.6 & & 2.6 &$\pm $& 0.6\\
\hline
$a_{q}$     &  &  $-$1.4 &$\pm$& 0.1 &  & $-$1.4 &$\pm$& 0.1\\
%\hline
%$\chi^{2}$/d.f. &             &           \\
\hline\hline
\end{tabular}
\hfil
\caption{Results from the NLO QCD fit to all
available published data compared to the results from a fit including also
projected data from HERA.
 The parameters $\eta_{NS}^{\rm p,n}$ are fixed 
by the octet hyperon $\beta$ decay constants: $g_A=F+D=1.2601\pm0.0025~\cite{ga}$ and $F/D  = 0.575\pm0.016~\cite{fd}$, ($a_8=3F-D$). The parameter $\eta_g$ is
 the first moment of the gluon distribution. All values of the parameters are given at $Q^{2}=1$ GeV$^{2}$. Errors reflect the statistical and systematic
errors on the data.}
\label{tab-results}
\end{table}
%%%
%%%  Figure 2
%%%
%The results are shown in Fig.~\ref{figbfrpd}. 
The first moment of the gluon distribution resulting from this fit is:
$\eta_g = 0.9 \pm 0.3 ({\rm exp}) \pm 1.0 ({\rm theory})$ at $Q^2=1~{\rm GeV}^2$. Similar results were obtained by an analysis performed by 
Altarelli et al.~\cite{ABFR}. The experimental error depends on the 
statistical and systematic uncertainty in the measured data points, while 
the theoretical uncertainty comes mainly from the choice of the factorization
and renormalization scales ($\delta \eta_g = \pm 0.6$), the
 initial parton distribution ($\delta \eta_g = \pm 0.5$) and the 
uncertainty in the value of $\alpha_{s}$ ($\delta \eta_g = \pm 0.3$).
 The polarized gluon distribution
 is deduced mostly 
from the observed scaling violations in the intermediate and low-$x$ regions.
Clearly the uncertainty on the size of the polarized gluon distribution is
large.  The existing  data poorly constrain the low-$x$  behavior 
of the various parton distributions  and 
hence the values of $\alpha_{\Sigma},\alpha_{g}$, 
and $\alpha_{NS}$. Moreover  
different functional forms for the initial parton distributions
result in  different predictions for the low-$x$
 behavior of $g_{1}(x)$ as well as different values of the
first moment of the polarized parton distribution~\cite{ABFR}, although the quality of 
the fit remains practically the same.

\begin{figure}[htb]
\epsfxsize=12cm
\epsfysize=12cm
\hfil
\epsffile[20 20 540 540]{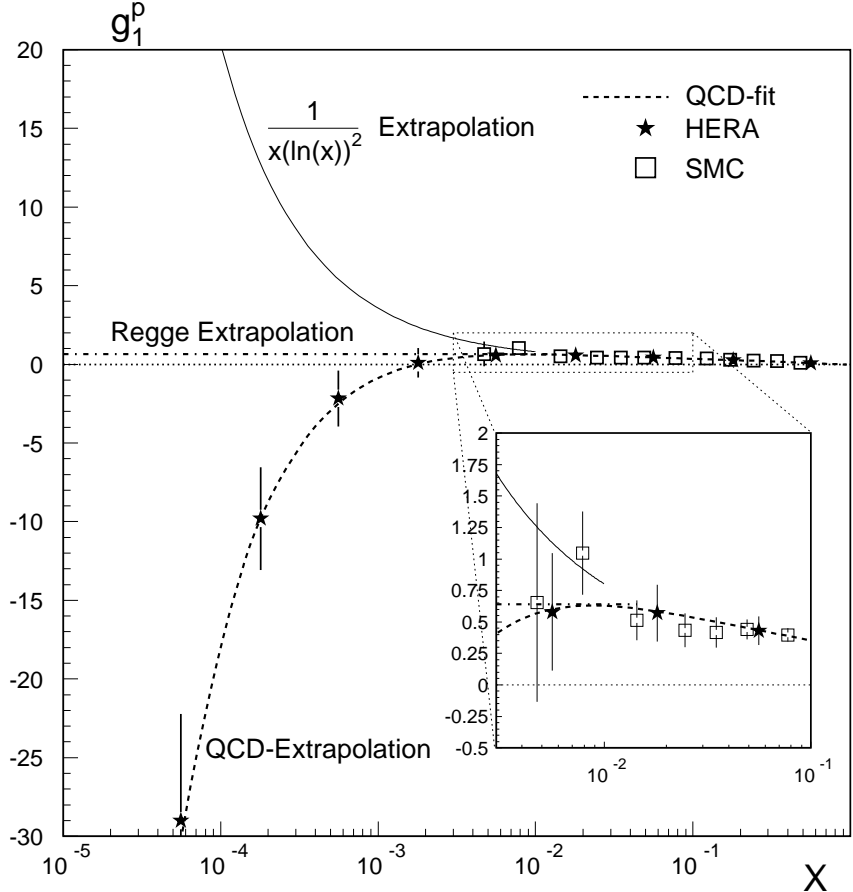}
\hfil
\caption{The  statistical uncertainty on the 
        structure function $g_{1}^{\rm p}$
         measurable at HERA for $Q^{2}=10$~GeV$^{2}$ with an integrated 
        luminosity ${\cal L}=500$~pb$^{-1}$ and a comparison of various 
  low-$x$ behaviors of $g_1^{\rm p}$. Also shown are the SMC measurements.
 Starting from the measured values of 
        $g_{1}^{\rm p}$ by SMC, the possible values for $g_{1}^{\rm p}$ at
        low-$x$ are shown for three scenarios: valence-like Regge behavior
        $g_{1}(x) \propto x^{-\alpha }$ with $\alpha = 0$, a strong powerlike
        positive rise $g_1(x)\propto
  1/(x \cdot ({\rm ln}x)^2)$, and the pQCD
        prediction based on the present best fit.
        A detailed comparison in the low-$x$ region of the SMC data is 
        shown in the inset.}
\label{lowx}
\end{figure}

The behavior of $g_1$ at low $x$ is still unknown.
Unpolarized structure function
data measured at HERA showed that $F_2$ rises at low $x$ which is in 
agreement with pQCD~\cite{rdbdr}, and for $Q^2 > 1~{\rm GeV}^2$
 does not follow a Regge behavior. 
For polarized DIS the 
low-$x$ information is crucial to obtain a good
determination of the moments of $g_1$~\cite{SMCp96,E154qcd,ABFR}.
In the past the extrapolation 
of $g_1$ from the measured region down to $x=0$, required to obtain its
first moment, was done
assuming a Regge behavior of the structure function, which 
implies~\cite{heim} $g_1\propto x^{-\alpha}$ as $x\to 0$ with 
$0\le \alpha \le 0.5$, i.e. a valence-like behavior of $g_1$. This 
behavior disagrees with that  predicted by  QCD. A comparison between
the Regge and QCD extrapolations from a recent analysis by the 
SMC~\cite{SMCp96} is shown in Fig.~\ref{lowx}.
%%%
%%%  Figure 3
%%%
The low-$x$ behavior has significant consequences on the first moment of $g_1$.
While the integral in the measured $x$-range of $0.003<x<0.7$ 
amounts to $0.139\pm 0.006({\rm stat.}) \pm 0.008({\rm syst.}) \pm
0.006 ({\rm evol.})$ at $Q^2=10~{\rm GeV}^2$ (the last error is due to
the uncertainty in the $Q^2$ evolution), the contribution from the
 unmeasured low-$x$ range 
 amounts to  0.002 or $-0.011$
depending on whether a Regge or a
pQCD prescription is used to extrapolate into the low-$x$ region. 
This provides  the  largest uncertainty on the first moment of $g_1$.
Hence, an experimental measurement of the 
low-$x$ behavior of $g_1$  would  lead to significant insight on 
the validity of  QCD.
% both within and beyond perturbation theory.

%%%%%%%%%%%%%%%%%%%%%%%%%%%%%%%%%%%%%%%%%%%%%%%%%%%%%%%%%%%%%%%%%%%%%
%%%%%%%%%%%%%   SECTION 3
%%%%%%%%%%%%%%%%%%%%%%%%%%%%%%%%%%%%%%%%%%%%%%%%%%%%%%%%%%%%%%%%%%%%%

\section{Measurement of {\boldmath  $g_{1}^{\rm p}(x,Q^{2})$} at HERA and its impact}
Presently all measurements of polarized structure functions are made
by  deep inelastic lepton-nucleon scattering on fixed targets. 
To evaluate the impact of measurements in the extended kinematic range accessible at HERA, a detailed study has been performed 
assuming an integrated luminosity 
of ${\cal L}=500$~pb$^{-1}$. After the planned luminosity upgrade of HERA~\cite{bartel}
this would be achievable in about three years of data-taking. 
Beam polarizations were assumed to be $0.7$  for electrons and protons.
Kinematic cuts were applied 
based on  standard analyses of data from the present 
detectors at the HERA collider~\cite{F2H1,F2ZEUS}: 
\begin{itemize}
\item
 the angle\footnote{Angles are defined with
 respect to the direction of the proton beam.} of the scattered 
electron, $\theta_e'$, is
required to be $< 177^{\rm o}$;
\item
 the inelasticity $y$ is limited to 
$0.01 < y < 0.9$;
\item 
the energy of the 
scattered electron, $E_e'$, is required to be $>5$~GeV;
\item
data were taken with $Q^2 > 1$~GeV$^2$.
\end{itemize}
A new parametrization~\cite{SMCp96} for the unpolarized structure 
function $F_2$,
which includes recent data from NMC, E665, H1 and ZEUS was used.
The SLAC parametrization~\cite{RSLAC} was 
used for the ratio $R$ of longitudinal and
transverse $\gamma^* p$ cross sections.
Results for the expected number of events 
for different $x-Q^2$ bins are  listed in Table~2, together with the 
average $y$, the depolarization factor $D=(y(2-y))/(y^2+2(1-y)(1+R))$, 
and the expected statistical uncertainty on the 
measured asymmetry $\delta A_{m}$. 
\begin{table}[htb]
\hfil
\begin{tabular}{|c|c|c|c|c|c|}\hline\hline
$x$ & $Q^{2}$ GeV$^{2}$ & $y$ & $D$ & $N_{total}$  & $\delta A_{m}$ \\
\hline\hline
$5.6\times 10^{-5}$ & $1.8$ & $0.40$ & $0.47$ & $3.8\times 10^{6}$ & $5.1\times 10^{-4}$ \\
\hline
$1.8\times 10^{-4}$ & $1.8$ & $0.13$ & $0.13$ & $6.9\times 10^{6}$ & $3.8\times 10^{-4}$ \\
                    & $5.6$ & $0.40$ & $0.46$ & $2.8\times 10^{6}$ & $6.0\times 10^{-4}$ \\
\hline
$5.6\times 10^{-4}$ & $1.8$ & $0.04$ & $0.04$ &$1.0\times 10^{7}$ & $3.1\times 10^{-4}$ \\
                    & $5.6$ & $0.13$ & $0.13$ &$4.9\times 10^{6}$ & $4.5\times 10^{-4}$ \\
                    & $1.8\times 10^{1}$ & $0.40$ & $0.47$ & $1.4\times 10^{6}$ & $8.6\times 10^{-4}$ \\
\hline
$1.8\times 10^{-3}$ & $1.8$ & $0.01$ & $0.01$ & $1.4\times 10^{7}$ & $2.7\times 10^{-4}$ \\
                    & $5.6$ & $0.04$ & $0.04$ & $7.3\times 10^{6}$ & $3.7\times 10^{-4}$ \\
                    & $1.8\times 10^{1}$ & $0.13$ & $0.13$ & $2.4\times 10^{6}$ & $6.4\times 10^{-4}$ \\
                    & $5.6\times 10^{1}$ & $0.40$ & $0.47$ & $6.9\times 10^{5}$ & $1.2\times 10^{-3}$ \\
\hline
$5.6\times 10^{-3}$ & $5.6$ & $0.01$ & $0.01$ & $9.2\times 10^{6}$ & $3.3\times 10^{-4}$ \\
                    & $1.8\times 10^{1}$ & $0.04$ & $0.04$ & $3.6\times 10^{6}$ & $5.3\times 10^{-4}$ \\
                    & $5.6\times 10^{1}$ & $0.13$ & $0.13$ & $1.2\times 10^{6}$ & $9.3\times 10^{-4}$ \\
                    & $1.8\times 10^{2}$ & $0.40$ & $0.47$ & $3.1\times 10^{5}$ & $1.8\times 10^{-3}$ \\
\hline
$1.8\times 10^{-2}$ & $1.8\times 10^{1}$ & $0.01$ & $0.01$ & $4.1\times 10^{6}$ & $4.9\times 10^{-4}$ \\
                    & $5.6\times 10^{1}$ & $0.04$ & $0.04$ & $1.5\times 10^{6}$ & $8.3\times 10^{-4}$ \\
                    & $1.8\times 10^{2}$ & $0.12$ & $0.13$ & $5.1\times 10^{5}$ & $1.4\times 10^{-3}$ \\
                    & $5.6\times 10^{2}$ & $0.40$ & $0.47$ & $1.3\times 10^{5}$ & $2.8\times 10^{-3}$ \\
\hline
$5.6\times 10^{-2}$ & $5.6\times 10^{1}$ & $0.01$ & $0.01$ & $1.4\times 10^{6}$ & $8.5\times 10^{-4}$ \\
                    & $1.8\times 10^{2}$ & $0.04$ & $0.04$ & $4.4\times 10^{5}$ & $1.5\times 10^{-3}$ \\
                    & $5.6\times 10^{2}$ & $0.12$ & $0.13$ & $1.4\times 10^{5}$ & $2.7\times 10^{-3}$ \\
                    & $1.8\times 10^{3}$ & $0.40$ & $0.47$ & $3.7\times 10^{4}$ & $5.2\times 10^{-3}$ \\
\hline
$1.8\times 10^{-1}$ & $1.8\times 10^{2}$ & $0.01$ & $0.01$ & $3.1\times 10^{5}$ & $1.8\times 10^{-3}$\\
                    & $5.6\times 10^{2}$ & $0.04$ & $0.04$ & $9.8\times 10^{4}$ & $3.2\times 10^{-3}$\\
                    & $1.8\times 10^{3}$ & $0.13$ & $0.13$ & $2.8\times 10^{4}$ & $6.0\times 10^{-3}$\\
                    & $5.6\times 10^{3}$ & $0.40$ & $0.47$ & $6.9\times 10^{3}$ & $1.2\times 10^{-2}$\\
\hline
$5.6\times 10^{-1}$ & $5.6\times 10^{2}$ & $0.01$ & $0.01$ & $1.4\times 10^{4}$ & $8.4\times 10^{-3}$\\
                    & $1.8\times 10^{3}$ & $0.04$ & $0.04$ & $3.9\times 10^{3}$ & $1.6\times 10^{-2}$\\
                    & $5.6\times 10^{3}$ & $0.13$ & $0.13$ & $9.8\times 10^{2}$ & $3.2\times 10^{-2}$\\
                    & $1.8\times 10^{4}$ & $0.40$ & $0.47$ & $2.1\times 10^{2}$ & $6.9\times 10^{-2}$\\
\hline\hline
\end{tabular}
\hfil
\caption{ The kinematic variables $x,Q^{2},y,$ and $D$, with the
number of events expected $N_{\rm total}$, and the statistical uncertainty in
the measured asymmetry $\delta A_{\rm m}$ assuming
an integrated luminosity ${\cal L} =500~{\rm pb}^{-1}$ and proton and electron
beam polarizations $P_{\rm p}=P_{\rm e}=0.7$.
}
\end{table}
Standard deep inelastic 
scattering formulae~\cite{g1old,yaledis} were used to calculate the 
kinematics and statistical uncertainties on the measured asymmetries. 
Using the best-fit parton distributions obtained from the QCD analysis
of the data discussed in Sect.~2, predictions
%Based on the pQCD prediction from the fit discussed in Sect.~2 predictions 
were made 
for  $g_1(x,Q^2)$ values and hence for cross section asymmetries 
in the HERA range. Systematic uncertainties were not included and will
be discussed at the end of this section.
As shown in Fig.~\ref{xq2_hera}, measurements of $g_1$ at HERA will 
extend the $x$ region down to $x=5.6\times 10^{-5}$. In Fig.~\ref{lowx} 
we show the expected accuracy of the determination of $g_1$ at HERA with 
the maximal extent of variation in the low-$x$ behavior  compatible 
with the requirement of integrability of $g_1$ (which implies that at 
small $x$ $g_1$ can rise at most as $1/(x \ln^\alpha x)$ with $\alpha>1$).

For the HERA measurements only one data point is shown at each $x$ value, 
chosen to be the point with the smallest statistical errors which 
corresponds to the largest depolarization factor (see Table~2). 
The figure also shows that a polarized HERA can clearly distinguish 
between the different prescriptions for the $g_{1}^{\rm p}$ behaviors in
the low-$x$ region.
%%%
%%%  Figure 4
%%%
\begin{figure}[htb]
\epsfxsize=19cm
\hfil
\epsffile[20 20 600 500]{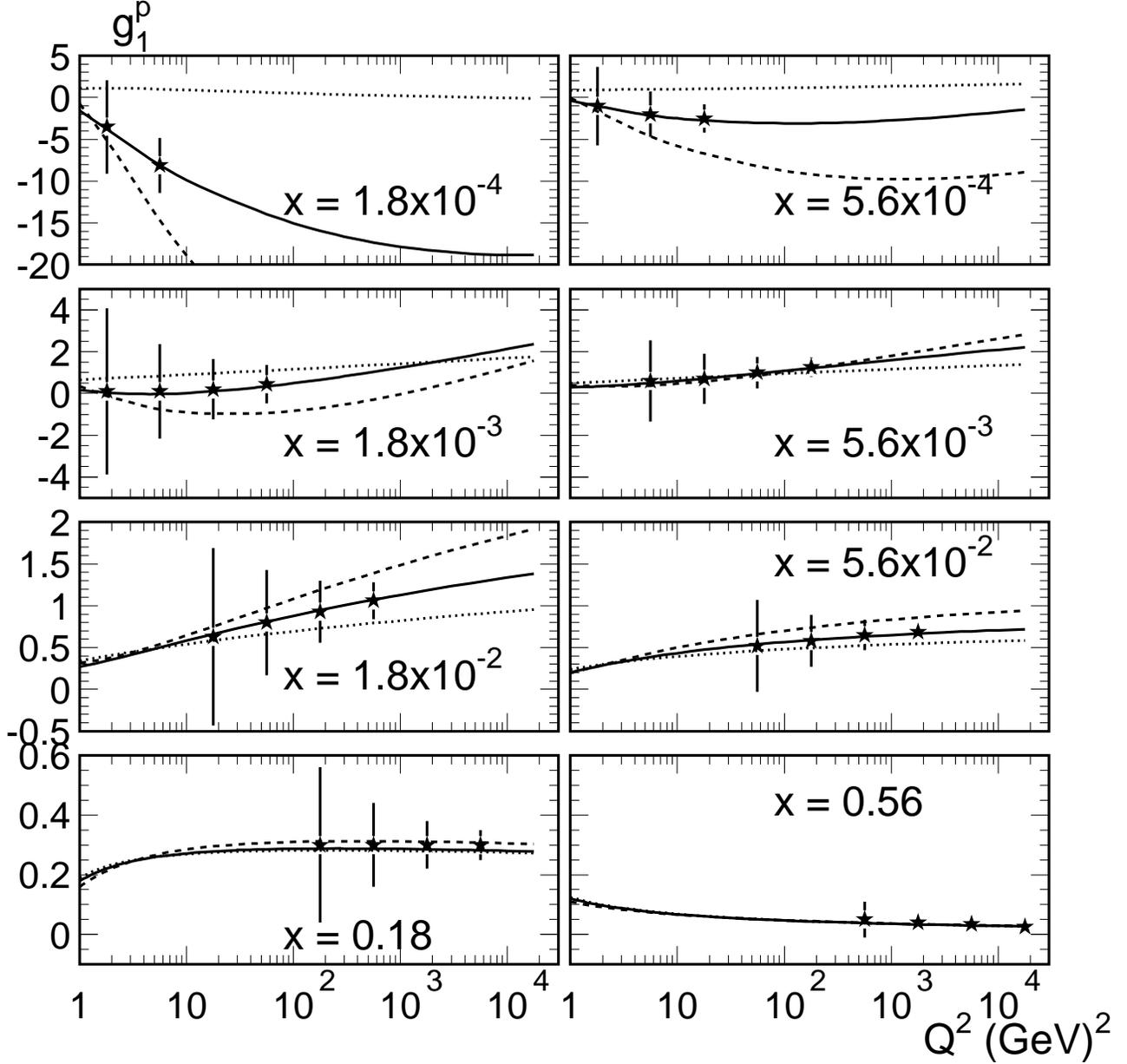}
  \caption{ Predicted values for $g_{1}^{p}$ from NLO fits together with
             estimated
             statistical uncertainties for a future polarized DIS experiment
             with an integrated luminosity ${\cal L}$=500~pb$^{-1}$
and with beam polarizations $P_{\rm e}=P_{\rm p} =0.7$.
             The solid lines are predictions based on fits to present
 data extended into the HERA kinematic range, while
             the dotted and dashed lines are the predictions
with a maximal and minimal gluon, respectively (see text).}
\label{scaling}
\end{figure}
 
The measurable points with their estimated errors  for HERA experiments 
are shown in Fig.~\ref{scaling} for eight accessible $x$-bins 
(out of the nine bins listed in Table~2).
%These are the bins were significant scaling violations can be observed. 
They are predicted on the basis of 
the NLO fit of Table~1, represented by the solid line.
We also consider two limiting scenarios: a) the first moment of 
the gluon distributions is fixed to be 0 at Q$^{2}$ = 1 GeV$^2$ (minimal 
gluon: dashed lines in Fig.~\ref{scaling}) and  b) the first
moment of the singlet quark density was fixed to $\eta_q = a_{8}$ at 
the same reference scale (maximal gluon: dotted lines in Fig.~\ref{scaling}).
Of course still wider deviations from these fits are possible since the 
low-$x$ behavior of the current best fit is only very loosely constrained 
due to the lack of direct experimental information at low $x$. 
Furthermore, 
higher order corrections beyond NLO may turn out to be important at very 
low $x$~\cite{NNLO1,NNLO2}.

The polarized gluon distribution is deduced from the scaling violations of  
$g_1$. Therefore  a reasonably wide kinematic coverage is required in order 
to achieve such a determination with satisfactory accuracy.
 In particular, most of the present-day parametrizations of polarized
parton distributions~\cite{BFRa,GRSV,SG}  indicate
that $\Delta g(x,Q^2)$ is peaked in the unmeasured low-$x$ region.
Therefore data in this region 
with a large span in $Q^2$, such as those obtainable at HERA, would
substantially improve the determination of $\Delta g(x,Q^2)$.

To assess the impact of these data we have repeated the QCD analysis 
described in Sect.~\ref{sec-g1stat} and included
the projected HERA data. The values of $g_{1}$ at HERA predicted by the QCD 
analysis were randomized within their estimated statistical errors
by assuming a Gaussian distribution.
The values of the best-fit parameters (Table 1)
 of course remain the same within
the errors, but  the calculated statistical errors on the projected 
measurements, used in the QCD analysis 
provide an estimate on the reduction of the experimental 
uncertainty in the fitted
 parameters.
In Table~\ref{tab-results} we
compare the result of the fit with the projected HERA data to the analysis of
 present-day data only. 
Note in particular the sizable improvement in 
the determination of the first moment of $\Delta g$, 
from the present error $\delta(\eta_g) = \pm 0.3$ to $\pm 0.2$ 
with HERA data.
A similar improvement is observed for the uncertainty in $\alpha_g$, which 
is determined by the low-$x$ data.
The theoretical uncertainty on the first moment of $\Delta g$ as of today
is large. The major contributing sources are factorization and renormalization
scales, the initial parton distributions and the uncertainty in the value 
of $\alpha_s$. At least the first two are direct consequences of the fact
that $g_1$ structure function measurements are available from a rather small
$x$ and $Q^2$-range, namely $0.003<x<0.8$ and $1<Q^2<100$~GeV$^2$. With HERA
the measured $x$ and $Q^2$-range would increase substantially, 
$10^{-5} < x < 0.8$ and $1<Q^2<10^4$~GeV$^2$.
Consequently the principal sources of theoretical uncertainty are
expected to reduce significantly. 
Out of all the systematic uncertainties we tried to study the most
dominant one, namely the factorization and renormalization scale 
uncertainty. This uncertainty is related to the 
lack of availability of higher than next-to-leading order corrections
and is usually estimated by changing the scale factors, $k_1,k_2$ (see
Section~2), in both directions (up and down), e.g.\ by a factor 
of two~\cite{ABFR} or four~\cite{GRSV}.
The difference between
the central value  $\eta_g^{best}$ and the one obtained from the fit
 which gives the largest deviation
from $\eta_g^{best}$ is assigned as the uncertainty
on $\eta_g$ due to the uncertainty in scales, $\delta_{scale}(\eta_{g})$.

If we follow this procedure using the present day data, the $\delta_{\rm scale}
(\eta_g) \approx \pm 0.6$. In order to estimate the improvement one
can expect in this uncertainty from future HERA data, this procedure
was repeated with the projected data in Table~2. The QCD analysis fails
to converge with the scale factors changed by factors of 4 or even 2,
indicating that with HERA data, this uncertainty is constrained already.
We repeated this procedure further by changing the scale factors
by smaller amounts ($[ k_1, k_2]= 0.5,0.8,1.2,1.5$) with the requirement 
that the fit procedure should converge and result in fits
with comparable $\chi^{2}$-confidence level.
The largest difference in value 
of $\eta_g$ obtained from such a procedure compared to the one in 
our present best fit was $\delta_{\rm scale}^{\rm HERA} = \pm 0.15$.
This suggests a possible improvement in this source of uncertainty 
by a factor of $\approx 4$.   
This improvement is indicative of the reduction one could achieve 
in the total theoretical uncertainty in $\eta_{g}$ from  HERA data.

%Although explicit re-evaluation of this is difficult, 
%preliminary investigations indicate that an improvement
%of at least a factor of two on each source is possible.
%The theoretical error has not been re-evaluated but it is expected 
%to decrease as well, especially due to the inclusion of low-$x$ data
% which reduces the
%uncertainty in the extrapolation. First studies show that the new data
%constrain the possible choice of the renormalization and factorization 
%scales, which are among the dominant sources of the theoretical errors. 

Systematic errors are discussed in~\cite{g1old,abhay}.
The principal types
of systematic errors associated with spin dependent asymmetry measurements
are
normalization errors and false asymmetries.
The normalization errors are mainly due to uncertainties in the electron
and proton beam polarizations $P_{\rm e}$ and $P_{\rm p}$. 
These lead to a
change in the magnitude of the measured asymmetry by an amount
which is small compared to the statistical error, and hence they are important
primarily when evaluating the first moment of $g_{1}^{\rm p}(x)$. 

Measurements of the electron polarization $P_{\rm e}$
at HERA by Compton scattering from 
polarized laser light have achieved~\cite{herm,barberpol}
 a relative uncertainty of $5\%$,
and plans are being made to improve it to the 2\%
level which has already been
achieved at SLAC~\cite{SLD}.
Absolute measurement of the polarization of the high energy  proton
beam presents a new challenge and is presently under investigation at
RHIC and at HERA.
It is reasonable to expect that an accuracy of 5\% can be 
achieved~\cite{schuler,leader,PpFermi,RHICSPIN}. The polarization errors
of 2\% (5\%) translate directly into a relative error of  2\% (5\%) on the
structure function $g_1$.

The other type of systematic error, which is due to false asymmetries,
arises from time variations in detector
efficiencies between the measurements for the parallel
and antiparallel spin conditions, and the correlation of proton 
polarization with the beam intensity or the bunch crossing angle. 
With regard to the time variation
of detector efficiencies, the interaction region for proton and electron
bunches
should be the same for different proton spin orientations, thus
 keeping the detector
acceptance constant.  
In the polarized HERA collider we expect to run in a mode where a 
sequence of about 10 proton bunches will have a given polarization, 
and the next sequence will have the opposite polarization. 
Since 
the time interval between the successive bunch crossings is 96 ns,
the asymmetry measurements will be done within 2 $\mu s$
and for such a time scale the detector efficiencies should be constant.
A possible correlation between proton bunch intensity and polarization
can lead to a false asymmetry, which can be avoided by repeating the 
measurements after rotation of all the proton spins. 
Small contributions to false asymmetries may arise due to the correlations
with the proton spin direction of the micro-structure and polarization
distribution within a bunch which can also be eliminated by the
proton spin rotation. 
A further possible source of false 
asymmetries comes from  potential asymmetries of  
background contamination in the $ep$ sample, namely interactions of the beams
with  beam-gas. Present analyses at HERA have achieved data samples
with a purity of better than 99\%. Furthermore, residual asymmetries
caused by such background interactions can be efficiently monitored
by using so-called pilot bunches, i.e.\ bunches which pass the 
interaction region and have no collision partner in the other beam.
Therefore this source of systematics can be kept smaller than 10$^{-4}$.
Hence we believe that 
false asymmetries at polarized HERA can be controlled at the required
level of $10^{-4}$ or less. Similar requirements are necessary for the 
RHIC-SPIN
program and are being actively studied. More quantitative estimates can not
be made at this time, since they depend on the 
polarized collider parameters and operating conditions which are
being studied at DESY.

We conclude that introduction of systematic errors will not
change significantly our conclusions on the impact of future HERA
$g_1$ data. This conclusion also applies
  to radiative corrections, which have been
studied for the HERA kinematics in~\cite{bardin} and 
we checked, using the program  {\tt HECTOR 1.11}~\cite{bardin1},
that they are well under control. 

%%%%%%%%%%%%%%%%%%%%%%%%%%%%%%%%%%%%%%%%%%%%%%%%%%%%%%%%%%%%%%%%
%%%%%%   Section 4
%%%%%%%%%%%%%%%%%%%%%%%%%%%%%%%%%%%%%%%%%%%%%%%%%%%%%%%%%%%%%%%%
\section{A direct measurement of {\boldmath $\Delta g(x,Q^2)$} from di-jet 
events at HERA} 
From the discussion in Sect.~3 it appears that
the polarized gluon
distribution $\Delta g(x,Q^2)$ is not as well  constrained 
by the $Q^2$ dependence of the structure function as is the 
unpolarized gluon density.
In fact, while the first moment of the polarized gluon density can be
deduced from scaling violations (including future HERA data) within
20\%, the shape of $\Delta g(x)$ is not constrained.
Therefore measurements which are directly sensitive to
$\Delta g(x,Q^2)$ are of great 
interest and are expected to add important information to the
determination of not only the first moment of $\Delta g$ but also of
its shape.
It has been shown by H1 at HERA~\cite{HERAG} that in leading order (LO) 
the unpolarized gluon 
momentum distribution, $g(x)$, can be accessed in a direct way by
measuring the cross section of di-jet production in DIS in a region of $Q^2=
5-50~{\rm GeV}^2$ with both jets measured in the central detector (the
proton remnant jet is not counted), and for jets with a transverse energy
$E_T$ larger than 5 GeV.
In leading order two  processes contribute to the di-jet events, 
 Photon-Gluon Fusion
(PGF) and QCD-Compton scattering (QCDC), both leading to two jets back to back
in azimuth in the hadronic center-of-mass system.
\begin{figure}[htb]
        \epsfxsize=12cm
        \hfil
        \epsffile[20 20 420 180]{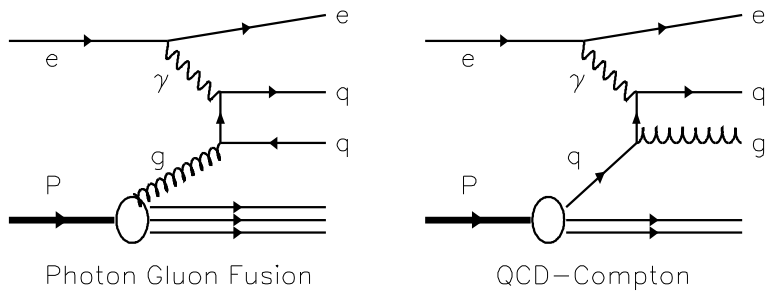}
        \hfil
        \caption{\label{fig:di-jets} Feynman diagrams for the
processes contributing to the di-jet cross section 
 at LO: the Photon-Gluon Fusion (PGF) process (left) and the QCD-Compton (QCDC)
process (right).}
\end{figure}
The Feynman diagrams for the two processes are
 shown in Fig.~\ref{fig:di-jets}. The PGF cross section is proportional
to $\alpha_s g(x)$.
The QCDC  cross section, however, is proportional
to $\alpha_s q(x)$ and constitutes the background.
The kinematic parameters to describe the PGF process are the momentum
fraction of the proton carried by the gluon $x_g$, the four-momentum transfer
$Q^2$ and the invariant
mass of the two jets $\hat{s}$. They are related to the Bjorken-$x$,
 by:
$$x_g=x (1+\frac{\hat{s}}{Q^2}).$$
For $\hat{s} > 100~{\rm GeV}^2$,
$x_g$ is larger than $x$ by about an order of magnitude
and the accessible range at
HERA is about $0.002 < x_g < 0.2$. 

The feasibility of using this kind of measurement in the case of
polarized lepton-nucleon 
scattering to extract $\Delta g(x,Q^2)$ from di-jet asymmetries
has been discussed  in~\cite{2jets-pub,2jets-proc-old}. 
A new study of such a  measurement at HERA  has been performed and is
detailed in~\cite{2jets-proc-new}. 
Recently  NLO corrections were calculated
 for the polarized jet
rates and found to be small~\cite{2jets-NLO}.
In this paper we summarize the main results and  use the 
projected data  to assess the  
sensitivity to the shape of $\Delta g(x,Q^2)$.
% to constrain the NLO fits to $g_1(x,Q^2)$ data.
In LO the total unpolarized di-jet cross section is the sum 
of the contributions from PGF processes  and QCDC
scattering, and can be written as:
\begin{equation}
\sigma_{\rm di-jet} =\sigma_{\rm di-jet}^{\rm PGF} +\sigma_{\rm di-jet}^{\rm QCDC} = A \,
\, g + B\,\,  q
\end{equation}
where $g$ and $q$ are the gluon and quark densities. $A$ and $B$ can
be calculated in perturbative QCD. Similarly, for the polarized case we
can write:
\begin{equation}
\Delta \sigma_{\rm di-jet} = \sigma_{\rm di-jet}^{\uparrow\downarrow} - \sigma_{\rm di-jet}^{\uparrow\uparrow} = a\,\,  \Delta g + b\,\, \Delta q 
\end{equation}
with $ \Delta g$ and $\Delta q$ being the polarized gluon and quark
densities. The
left arrow in the subscript denotes the polarization of the 
incoming lepton with respect to the direction of its momentum.
The right arrow stands for the polarization of the proton being anti-parallel 
or parallel to the polarization of the incoming lepton.
The di-jet asymmetry is  sensitive to $\Delta g/g$, especially
at low $x$ where the PGF cross section dominates.
\begin{equation}
A_{\rm di-jet} = \frac{\Delta \sigma_{\rm di-jet}}{2\, \sigma_{\rm di-jet}} = 
{\cal A}\,\,\frac{\Delta g}{g}\,\,\, \frac{\sigma_{\rm di-jet}^{\rm PGF}}
{2\, \sigma_{\rm di-jet}}
+ {\cal B}\,\,\frac{\Delta q}{q}\,\,\, \frac{1}{2}\, (1-\frac{\sigma_{\rm di-jet}^{\rm PGF}
}{\sigma_{\rm di-jet}}),
\label{eq-A}
\end{equation}
with ${\cal A} \equiv a/A$ and ${\cal B} \equiv b/B$. 
The experimentally accessible asymmetry $A_{\rm meas}$ is related
to $A_{\rm di-jet}$ by
\begin{equation}
\label{eq:Ameas}
A_{\rm meas} = 
\frac{N^{\uparrow \downarrow}-N^{\uparrow \uparrow}}
{N^{\uparrow \downarrow}+N^{\uparrow \uparrow}} 
= P_{\rm e} P_{\rm p} D A_{\rm di-jet}
\end{equation}
where $D$ is the depolarization
 factor and $A_{\rm meas}$
is the measured asymmetry.
The quantities $N^{\uparrow \downarrow}$ ($N^{\uparrow \uparrow}$)
are the total number of observed di-jet events 
($N^{\uparrow \downarrow}=N^{\uparrow \downarrow}_{\rm PGF}
+N^{\uparrow \downarrow}_{\rm QCDC}$) for 
antiparallel (parallel) proton and electron spins.
 
A Monte Carlo study was performed for a polarized HERA 
on the expected size and uncertainty
of the measured  di-jet asymmetries and 
the contributions from signal and background processes.
{\tt PEPSI 6.5}~\cite{PEPSI} was used, which is
 a full LO lepton-nucleon scattering Monte Carlo program
for unpolarized and polarized interactions, 
including fragmentation, and unpolarized parton
showers to simulate higher order effects.
On the parton level the results were cross checked using the Monte Carlo
cross section integration program~\cite{MEPJET}.
 Detector smearing
and acceptance effects were included based on the performance of the present
H1 detector.
As input for the unpolarized parton distributions we used 
the LO GRV parametrizations~\cite{GRV94} and for the polarized
 distributions various
published parametrizations, as well as the result for $\Delta g$ of the NLO fit
described in the previous section were used. All these distributions are
in agreement with the present data.
Monte Carlo events  were generated
for an integrated luminosity of 500 pb$^{-1}$ and
beam polarizations of $0.7$ for electron
 and proton beams.  
Jets were defined by the cone jet algorithm with a cone radius of 1 in 
the pseudorapidity $\eta$ and azimuthal angle $\phi$ plane. 
Di-jet events were selected by the following criteria: 
\begin{itemize}
\item
two jets with  $E_T > 5$~GeV;
\item
$5~{\rm GeV}^2 < Q^2 < 100~{\rm GeV}^2; $
\item
$0.3 < y < 0.85;$
\item
$\hat{s} > 400\, (100)\, {\rm GeV}^2.$
\end{itemize}
The invariant mass of the two jets  is reconstructed from their
 four-momenta $P_{1(2)}$:
$\hat{s} = (P_1 + P_2)^2$.
The cut $\hat{s} > 400~{\rm GeV}^2$
  was chosen to keep higher order corrections
small~\cite{2jets-NLO}. However, the data with $ 100 <  \hat{s} < 400~{\rm GeV}^2$ were also analyzed. The asymmetries were found to be as large 
as a few percent and are listed in 
 Table~\ref{2jets-asys}a.

\begin{table}[htb]
\hfil
%\begin{flushleft}
\hspace{-4.5cm}
\begin{tabular}{||c|c|c|c||}
\hline\hline
{\bf a)} & \multicolumn{3}{|c||}{$ 5 < Q^2 < 100~{\rm GeV}^2$} \\
$x_g$ & $A_{meas}$ & $A_{corr}$ & $\delta(A)$ \\
\hline \hline
0.002 & $-$0.003 & $-$0.003 & 0.005 \\
0.006 & $-$0.008 & $-$0.008 & 0.003  \\
0.014 & $-$0.013 & $-$0.013 & 0.004 \\
0.034 & $-$0.030 & $-$0.030 & 0.006\\
0.084 & $-$0.031 & $-$0.042 & 0.012 \\
0.207 & $-$0.023 & $-$0.047 & 0.027 \\
\hline\hline
\end{tabular}
%\hfil
%\end{flushleft}
\hfil
%\hspace{3cm}
\vspace{-4.69cm}
\begin{flushright}
\hspace{-5cm}
\begin{tabular}{||c|c|c||c|c||}
\hline\hline
{\bf b)} & \multicolumn{2}{|c||}{$ 2 < Q^2 < 10~{\rm GeV}^2$} &  \multicolumn{2}{|c||}{$
 10 < Q^2 < 100~{\rm GeV}^2$}\\
$x_g$ & $A_{corr}$ & $\delta(A)$  & $A_{corr}$ & $\delta(A)$ \\
\hline \hline
0.002 & $-$0.006 & 0.005 & $-$0.001 & 0.007\\
0.006 & $-$0.010 & 0.003 & $-$0.006 & 0.003\\
0.014 & $-$0.011 & 0.004 & $-$0.015 & 0.004\\
0.034 & $-$0.033 & 0.007 & $-$0.029 & 0.007\\
0.084 & $-$0.041 & 0.013 & $-$0.035 & 0.013\\
0.207 & $-$0.091 & 0.027 & $-$0.013 & 0.031\\
\hline\hline
\end{tabular}
\end{flushright}
\hfil
\caption{a) Expected measured asymmetries from di-jet events for six $x_g$-bins
with a polarized
 HERA and
 background corrected asymmetries with
statistical errors corresponding to an integrated luminosity of
500~pb$^{-1}$. b) The background corrected asymmetries for
two different $Q^2$ ranges.}
\label{2jets-asys}
\end{table}
The two lowest $x_g$-bins  in the table
 are obtained from  the lower $\hat{s}$ data.
In this table the asymmetry corrected for QCDC background
($A_{\rm corr} = A_{\rm meas} - \frac{N^{\uparrow \downarrow}_{\rm QCDC}-N^{\uparrow \uparrow}_{\rm QCDC}}
{N^{\uparrow \downarrow}+N^{\uparrow \uparrow}}$)
is also presented. Note that 
  there is a significant  QCDC contribution only for the
two highest $x_g$-bins. 
%Even with a relatively
%low luminosity of 200~pb$^{-1}$
% the statistical uncertainties  are small enough to 
%ensure the measurement of the di-jet asymmetries.
%%%
\begin{figure}
\epsfxsize=12cm
\hfil \epsffile[100 0 500 300]{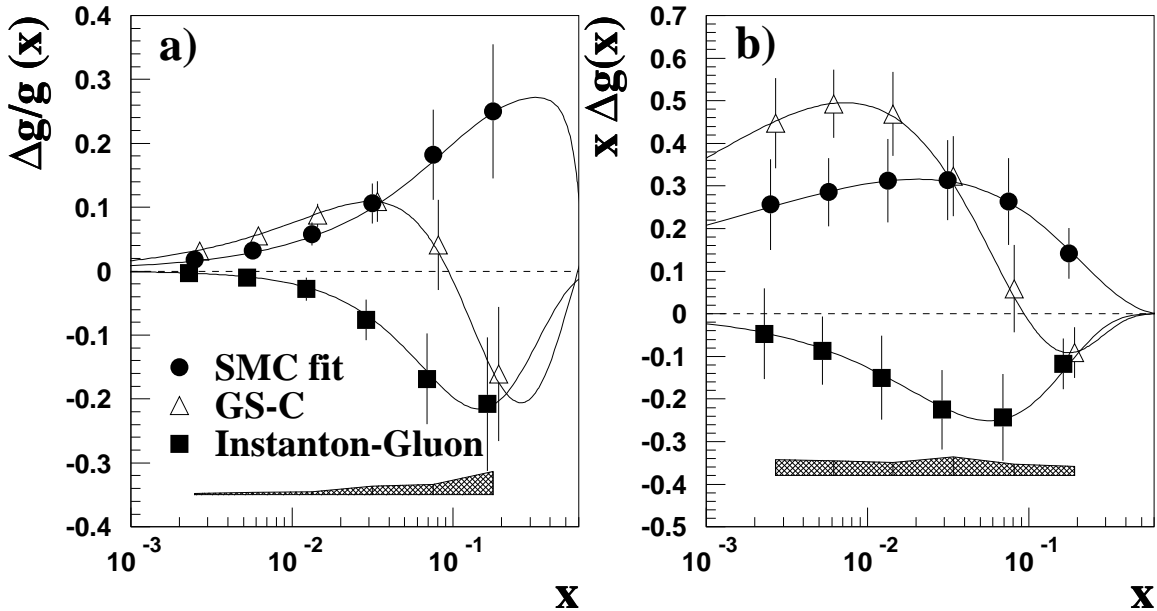} \hfil
\caption{ The sensitivity to
 $\Delta g(x)/g(x)$ (a) and $x \Delta g(x)$ (b) from a measurement of di-jet
 events with
 a polarized HERA at $Q^2=20~{\rm GeV}^2$
  assuming an integrated luminosity of 500~pb$^{-1}$.
The error bars show the expected statistical error. An estimate of systematic
uncertainties is reflected in the shaded error band. As input for
the parton distributions three different parametrizations were used (see text).} 
\label{2jets-deltag}
\end{figure}
Fig.~\ref{2jets-deltag} shows the projected values of $\Delta g(x)/g(x)$
and $x\Delta g(x)$ as a function of $x_g$ for ${\cal L}=500~{\rm pb}^{-1}$
for different $\Delta g(x,Q^2)$  distributions:
the polarized gluon density obtained from the NLO fit to $g_1$ data by the
SMC, the 
gluon set C by Gehrmann and Stirling~\cite{SG} and an instanton-induced gluon
proposed by Kochelev~\cite{Kochelev}. The results are presented 
 at $Q^2=20~{\rm GeV}^2$ which is very close to the average $Q^2$ of the
simulated data.
% The expected statistical errors show that di-jet
%events can
%distinguish between the different  $\Delta g$ distributions.

The shaded error band shows an estimate of the systematic
uncertainty considering the following error sources: an
uncertainty of 2\% on the calibration of the hadronic energy scale, 
the error on the total unpolarized di-jet cross section of 2\% and
on the unpolarized gluon density $g(x)$ of 
5\%~\cite{Klein}. The uncertainty on
the ratio of  the polarized and unpolarized quark densities, $\Delta q/q$, 
was considered to be 10\% and the error on the polarization measurements
for electron and proton beams was taken to be 5\%.
Both statistical and systematic errors  were calculated~\cite{2jets-proc-new}
using as input the gluon set A of Gehrmann and Stirling~\cite{SG},
which resembles the gluon distribution extracted from the NLO fit to $g_1$,
and were adjusted to the different parametrizations used here.
 
It can be seen that the measurement is sensitive to the different
gluon distributions, and that its shape can be determined with
six  bins in $x$. Using these projected measurements the statistical error
of the first moment of $\Delta g$ in the $x$-range of
0.0015--0.32 is: $\delta (\int_{0.0015}^{0.32} \Delta g(x) {\rm d}x) = 0.2$.
 
A further study was performed using additional information
on the $Q^2$ dependence.
For this study we extended the $Q^2$ range down  to $2~{\rm GeV}^2$
and divided the data into two bins with: $ 2 < Q^2 < 10~{\rm GeV}^2$
and $10 < Q^2 < 100~{\rm GeV}^2.$ The expected sensitivity
on $\Delta g(x,Q^2)$, reflected by the expected asymmetries and their errors,
 is shown in Table~\ref{2jets-asys}b.
The mean $Q^2$ values for the two bins are 4.5 and 30~GeV$^2$, respectively.
The effect of the $Q^2$ evolution of the gluon can be seen, but does not 
appear to be very large.

%%%%%%%%%%%%%%%%%%%%%%%%%%%%%%%%%%%%%%%%%%%%%%%%%%%%%%%%%%%%%%%%
%%%%%%   Section 5
%%%%%%%%%%%%%%%%%%%%%%%%%%%%%%%%%%%%%%%%%%%%%%%%%%%%%%%%%%%%%%%%
\section{Combined QCD analysis of {\boldmath $g_1$} data and {\boldmath 
 $\Delta g$} from di-jet asymmetries}

The measurement of $g_1(x,Q^2)$ and di-jet asymmetries
provide complementary information on the polarized
gluon distribution. 
As argued in Sections~3 and~4 
the first moment of the polarized gluon distribution
can be obtained from the QCD analysis in NLO of $g_1$ data
while di-jet asymmetries will 
provide a direct measurement of the polarized gluon
density $\Delta g(x,Q^2)$ in the range $0.0015<x<0.32$.
Hence, it is interesting
to study the impact  of  a combined
analysis of $g_1$ and di-jet asymmetry
data on the determination of $\Delta g$.

   In principle, such an analysis should be carried out using the
measured di-jet asymmetries and $g_1$ data in a self-consistent
procedure in NLO. Polarized quark and gluon densities should be fitted
simultaneously to both data sets, in order to reproduce the scaling violations
in $g_1$ and separate the PGF and QCDC contributions to the
jet asymmetry cross sections. The tools to
perform such  analysis are not available yet.
However, the effect of the combined analysis on the uncertainty
in $\Delta g$ can be assessed by including the values of $\Delta g(x,Q^2)$
obtained from the di-jet analysis in the NLO QCD fit  of the
$g_1$ data as an extra constraint.

\begin{figure}
\epsfxsize=12cm
\hfil 
\epsffile[0 260 530 530]{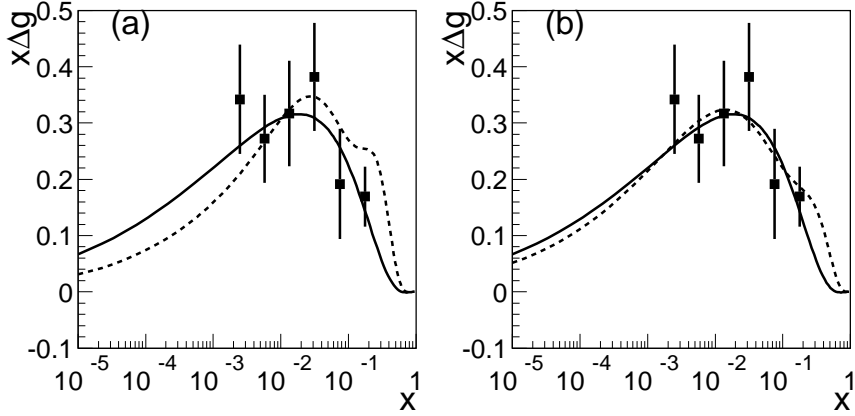} 
\hfil
%\vspace{-6cm}
%\hspace*{12cm}
\caption{\label{deltags} The polarized gluon distribtion at 
$Q^2 = 20$ GeV$^2$
obtained from a 14 parameter fit (Eq.\ 3), dashed line,
including only existing $g_1$ data (a), and including existing and
projected $g_1$ and $\Delta g(x)$ data (b). 
For comparison also the results of the
10 parameter fit to existing and projected data is shown, solid line,
as well as the corresponding $\Delta g$ values with their statistical errors,
as they could be obtained from a di-jet analysis.}
\end{figure}

The values of $\Delta g(x,Q^2)$ in the six $x$-bins, as obtained from the analysis
of di-jet asymmetries, were considered as data.  
Statistical errors were obtained from the Monte Carlo simulation discussed
in Sect.~4. The values used for $\Delta g(x,Q^2)$ were taken  from
 the best fit of present-day $g_1$ data described in Sect.~2.
Like the projected $g_1$ data
for the HERA kinematic region, they were randomized around the
calculated value within one standard deviation of the statistical
error.
As mentioned in Section 3, including the projected $g_1$ data
in the NLO analysis, reduced the error on the first moment of
$\Delta g$ by about 30\%, yielding $\eta_g = 0.9 \pm 0.2$,
down from $\delta(\eta_g) = 0.3$.
The QCD analysis of the present
and the projected $g_1$ data
was repeated including also 
$\Delta g(x,Q^2)$ values from the di-jet asymmetries in the fit.
While the best fit parameters are obviously about the same as those obtained
for the NLO fit of the $g_1$ data, the uncertainty on the first moment of the
polarized gluon distribution is reduced by 50\%, resulting in
$\eta_g=1.0 \pm 0.1$. The polarized gluon distribution from the best fit
of $g_1$ data, and projected HERA data, evolved to 20 GeV$^2$ is shown
in Fig.~\ref{deltags}  along with the projected data of $\Delta g(x,Q^2)$
 used in the fit.
%%%
%%%  Figure 5
%%%
\begin{table}[htb]
\hfil
\begin{tabular}{||c|c||}
\hline\hline
                                &      \\
Data used                       &   $\delta (\int \Delta g)$\\
                                &            \\
\hline
                                &  \\
QCD Anal. of published $g_{1}$ today&  $\pm 0.3$ \\
         (= ``A'')           & \\
\hline
                                &  \\
``A'' + Projected HERA $g_{1}^{\rm p}$  &  $\pm 0.2$            \\
                                &  \\
\hline
                                &   \\
``A'' + Projected HERA $\Delta g$-jets  &   $\pm 0.2$  \\
                                &   \\
\hline
                                &   \\
``A'' + Projected HERA $g_{1}^{\rm p}$ and $\Delta g$-jets
                                &     $\pm 0.1$        \\
                                &   \\
\hline\hline
\end{tabular}
\hfil
\caption{\label{tab-hera} The expected uncertainty in the determination of
        the first moment of the gluon distribution at $Q^{2} = 1$
         GeV$^2$ with inclusive and di-jet data  from
         polarized HERA with ${\cal L} = 500$ pb$^{-1}$.}
\end{table}
The effect of the di-jet data alone on the uncertainty in $\eta_g$
is similar to that of the additional $g_1$ data which will be measured
at HERA. The combined fit of present-day $g_1$ data with the $\Delta 
g(x,Q^2)$ from di-jet
asymmetries yields $\eta_g = 1.0 \pm 0.2$.
The results for the first moments of the polarized gluon distributions
obtained from the fit of $g_1$ data with projected $g_1$ and $\Delta g$ data
are summarized in Table~4.
%The improved determination of the gluon distribution in the combined fit
%is also reflected in the further
% reduced uncertainty in $\alpha_g$, which determines
%the behavior of $\Delta g(x,Q^2)$ at low $x$. The value obtained from the
%fit of the $g_1$ data is $\alpha_g = -0.5 \pm 0.3$. When the projected
%$g_1$ and $\Delta g(x,Q^2)$ data  are also considered in the fit we obtain
%$\alpha_g = -0.6 \pm 0.1$.

While the first moment of the polarized gluon distribution
can be deduced from scaling violations in $g_1$, the QCD analysis
does not constrain its shape. This can be demonstrated
by introducing an additional term, $b_f \sqrt{x}$ into the functional form
of the parton distributions,
\begin{equation}
\label{part_dist_a}
\Delta f(x,Q^{2}) =
N(\alpha_f,\beta_f,a_f)~\eta_f~x^{\alpha_f}(1-x)^{\beta_f}(1+a_f~x+ b_f~\sqrt
x).
\end{equation}
Using this form in the QCD analysis of the $g_1$ data
results in a fit with a node
in $\Delta g(x,Q^2)$, which is not allowed by the more restricted
form of Eq.\ 2.  The gluon distribution obtained from a fit with this
functional form is shown in
 Fig.~\ref{deltags} where it is also compared with the
gluon distribution obtained with the standard form (Eq.\ 2). While
  the first moment
of $\Delta g$ at $Q^2=1~$GeV$^2$ is 0.9 $\pm$ 0.3,
 the same value as 
obtained in the standard fit, the shapes of the gluon distributions are 
different.
%Obviously, this fit is not compatible
%with the $\Delta g(x)$ values obtained from the fit using the
%parton distributions of Eq.\ 2.

On the other hand, di-jet asymmetries provide good
information on the shape of the polarized gluon distribution in the measured
region.
The effect of the additional data on our knowledge of $\Delta g$ can be
demonstrated when  the projected values of $g_1$ and $\Delta g(x,Q^2)$, 
(obtained
from the standard, 10 parameter fit) are fitted with 14 parameters, as
discussed above.
As shown in Fig.~\ref{deltags}, once all the projected $g_1$ and di-jet
data are included in the fit, the polarized gluon distribution obtained from
the 14 parameter fit is essentially identical to that obtained from the 10
parameter fit
in the region where $\Delta g(x,Q^2)
$ data are introduced. Obviously, $\Delta g(x,Q^2)$
remains unconstrained in the high $x$ region
where no $\Delta g(x,Q^2)
$ data is added and where the $Q^2$ evolution is small.
One may thus conclude that the combined analysis will constrain the
functional form of the polarized gluon distribution in addition
to its first moment.

%%%%%%%%%%%%%%%%%%%%%%%%%%%%%%%%%%%%%%%%%%%%%%%%%%%%%%%%
%\newpage
\section{Conclusions}
Inclusive polarized DIS measurements at HERA with polarized protons
and with a high integrated luminosity 
will  yield
significant and unique new information on $g_{1}^{\rm p}(x,Q^{2})$ over a much
extended range in $x$ and $Q^{2}$, which is of great theoretical interest.
Statistical errors on the points will dominate. 
 For such measurements the false asymmetries
should be considerably smaller than the true asymmetries to be measured, and
normalization errors should be controlled to be better than $10\%$. 

The present uncertainties in the determination of the first moment of 
the polarized gluon distribution are large. They will be considerably
reduced by the pQCD analysis of the $g_{1}$ structure function
data from HERA. However
we have shown that even further reduction in the uncertainty can be achieved
if polarized gluon information coming from the di-jet events is used 
in the analysis in conjunction with the $g_{1}$ data.

The fact that HERA is an operating facility with a plan for substantial
increase in luminosity and with two major operating detectors, as well
as a polarized electron beam, means that only the high energy polarized 
proton beam needs to be developed. The efforts to achieve this goal seem well 
justified and data from 
such experiments at HERA would provide unique and complementary information 
to that from other presently proposed experiments.

%\newpage
\hspace{-0.6cm}{\bf Acknowledgements:} 
We thank G. Altarelli, R. Ball, D. Bardin, J. Ellis, S. Forte and G. Ridolfi
for many stimulating and helpful discussions.
This work is supported in part by the U.S. Department of Energy 
and by the Israel Science Foundation of the Israeli Academy of Sciences.

%%% The bibliography %%%%

%%%%%%%%%%%%%% TABLES %%%%%%%%%%%%%%%%%%%%%%%%%%%%%%%%%%%%%%%%%%%%%%%%%%%%%%%%%%%%

%%%%%%%%%%%%% Figures %%%%%%%%%%%%%%%%%%%%%%%%%%%%%%%%%%%%%%%%%%%%%%%%%%%%%%%%%%%
%\newpage

%%%%%%%%%%%%%%%%%%%%%%%%%%%%%%%%%%%%%%%%%%%%%%%%%%%%%%%%%%%%%%%%%%%%%%%%%%%%%%%%
%%%%%%%%%%%%%%%%%%%%%%%%%%%%%%%%%%%%%%%%%%%%%%%%%%%%%%%%%%%%%%%%%%%%%%%%%%%%%%%%
%%%%%
%%%%%                     END OF THE DOCUMENT
\end{document}